\documentclass[iop]{emulateapj}
\usepackage{arydshln}
\slugcomment{{Accepted for publication in the Astrophysical Journal Letters}}
\usepackage{epsfig, natbib}
\usepackage{subfigure}
\usepackage{arydshln}
\def\arcsec{\hbox{$^{\prime\prime}$}}

\usepackage{xcolor}

\begin{document}
\title{\textit{Chandra} observations of the newly discovered magnetar Swift J1818.0--1607}
\author{Harsha Blumer\altaffilmark{1,2} and Samar Safi-Harb\altaffilmark{3}}
\altaffiltext{1}{Department of Physics and Astronomy, West Virginia University, Morgantown, WV 26506, USA; harsha.blumer@mail.wvu.edu}
\altaffiltext{2}{Center for Gravitational Waves and Cosmology, West Virginia University, Chestnut Ridge Research Building, Morgantown, WV 26505, USA}
\altaffiltext{3}{Department of Physics and Astronomy, University of Manitoba, Winnipeg, MB R3T 2N2, Canada; samar.safi-harb@umanitoba.ca}

\begin{abstract}

Swift J1818.0--1607 is a new radio-loud magnetar discovered by the {\it Swift} Burst Alert Telescope on 2020 March 12. It has a magnetic field $B$$\sim$2.5$\times$10$^{14}$~G, spin-down luminosity $\dot{E}$$\sim$7.2$\times$10$^{35}$ ergs~s$^{-1}$, and characteristic age $\tau_c$$\sim$470~yr. Here we report on the {\it Chandra} observations of Swift J1818.0--1607, which allowed for a high resolution imaging and spectroscopic study of the magnetar and its environment. The 1-10 keV spectrum of the magnetar is best described by a single blackbody model with a temperature of 1.2$\pm$0.1~keV and an unabsorbed flux of 1.9$_{-0.3}^{+0.4}\times$10$^{-11}$~ergs~cm$^{-2}$~s$^{-1}$. This implies an X-ray luminosity of 9.6$_{-1.5}^{+2.0}\times$10$^{34}$~$d_{6.5}^2$~ergs~s$^{-1}$ and efficiency of $L_{X}/\dot{E}$$\sim$0.13~$d_{6.5}^2$ at a distance of 6.5~kpc. The {\it Chandra} image also shows faint diffuse emission out to $\geq$10$\arcsec$ from the magnetar, with its spectrum adequately described by a power law with a photon index of 2.0$\pm$0.5 and a luminosity of $\sim$8.1$\times$10$^{33}$~$d_{6.5}^2$~ergs~s$^{-1}$. The extended emission is likely dominated by a dust scattering halo and future observations of the source in quiescence will reveal any underlying compact wind nebula. We conclude that Swift J1818.0--1607 is a transient source showing properties between high-$B$ pulsars and magnetars, and could be powered at least partly by its high spin-down similar to the rotation-powered pulsars. 

\end{abstract}


\section{Introduction}
\label{1}

Magnetars are young neutron stars with extremely strong magnetic fields ($B$), often surpassing $\sim$10$^{14}$~G, and believed to be powered mainly by magnetic energy dissipation. They emit short ($\sim$0.1 s), bright hard X-ray bursts over the course of days to months, accompanied by changes to its spectral and temporal properties or radio detection (Kaspi \& Beloborodov 2017). Several observational results in the last few years have demonstrated that the magnetar family extends beyond the canonical aforementioned observational properties.  This includes discovery of magnetar-like bursts from the rotation-powered pulsars (PSRs) J1846--0258  (Gavriil et al. 2008; Kumar \& Safi-Harb 2008) and J1119--6127 (Younes et al. 2016; Gogus et al. 2016; Archibald et al. 2016), central compact objects (Rea et al. 2016), and low $B$ magnetars (e.g., Rea et al. 2010). These results present a remarkable set of observational properties of the magnetar population and a testing ground for different emission mechanism models in these sources. 

On 2020 March 12, the {\it Swift} Burst Alert Telescope (BAT) on board the Neil Gehrels Swift Observatory triggered on a short ($\sim$0.1 s), soft burst indicating the discovery of a new magnetar (Evans et al. 2020) and the {\it Swift} X-ray Telescope (XRT) observations revealed an uncatalogued X-ray source, Swift J1818.0--1607 (hereafter, J1818). This was followed by NICER observations which led to the detection of a coherent periodicity at 0.733417(4) Hz with significance exceeding 5-sigma (Enoto et al. 2020). The short burst detected with {\it Swift} together with NICER observations suggest that J1818 is a new Galactic magnetar with a spin period of 1.36 s, which is the shortest among the known magnetars, but longer than the period of high-$B$ field rotation-powered pulsars with magnetar-like activity (Enoto et al. 2020). 

\begin{figure*}[ht]
\includegraphics[width=\textwidth]{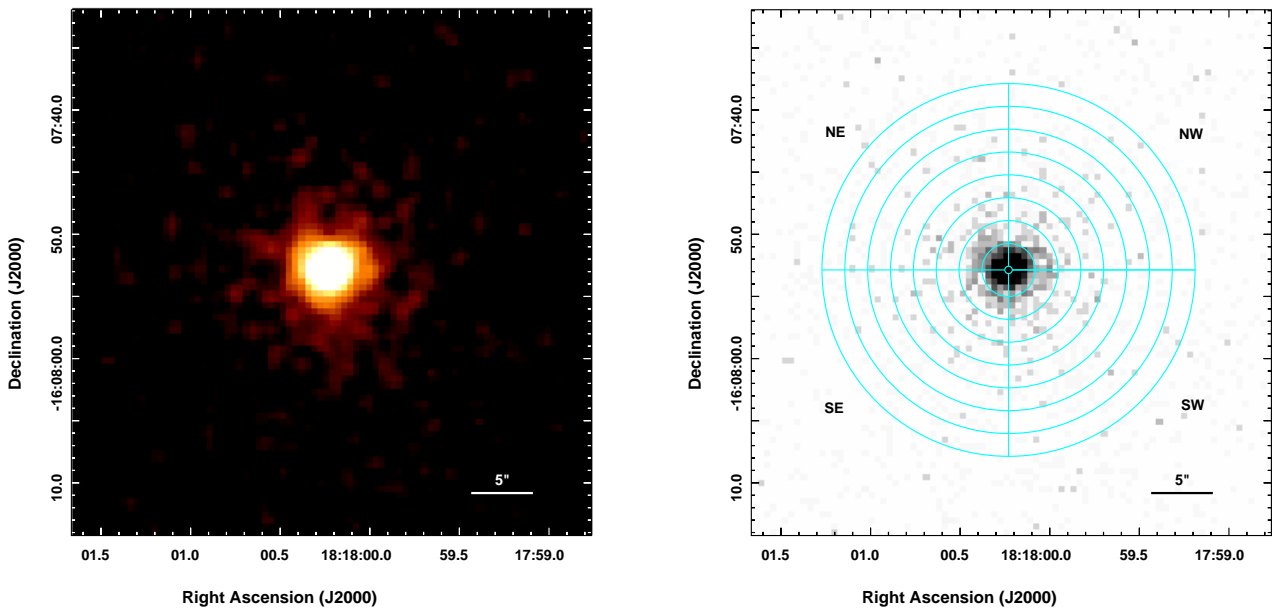}
\caption{Images of J1818 and surrounding diffuse emission in the 0.5--7 keV energy band. The images are exposure-corrected and shown in logarithmic scale. The smoothed (left) and unsmoothed (right) images are in units of photons~cm$^{-2}$~s$^{-1}$~arcsec$^{-2}$ and counts pixel$^{-1}$, respectively. Overlaid on the unsmoothed image are annular regions divided into four quadrants (shown in cyan), oriented to the north (N), east (E), south (S), and west (W), for radial profile analysis of the diffuse emission. North is up and East is to the left. See Section~3 for details. }
\end{figure*}

Radio follow-up observations by the 100-m Effelsberg radio telescope in a band centered on 1.37 GHz  identified J1818 as the fifth radio-loud magnetar with a dispersion measure (DM) of 706$\pm$4 pc cm$^{-3}$ (Karuppusamy et al. 2020) and provided a first measurement of the spin period derivative of 8.2$\times$10$^{-11}$~s s$^{-1}$ (Champion et al. 2020). Simultaneous observations were performed with {\it XMM-Newton} and {\it NuSTAR} three days after the magnetar burst, where the source spectrum was described by an absorbed blackbody (BB) with interstellar absorption $N_H$=(1.22$\pm$0.03)$\times$10$^{23}$~cm$^{-2}$ and temperature $kT$=1.19$\pm$0.02 keV plus a powerlaw (PL) with photon index $\Gamma$=0.1$\pm$1.2 in the 0.3--20 keV energy band (Esposito et al. 2020). The X-ray timing analysis performed with NICER suggested a dipolar magnetic field $B$$\sim$2.5$\times$10$^{14}$~G, spin-down luminosity $\dot{E}$$\sim$7.2$\times$10$^{35}$ ergs~s$^{-1}$, and characteristic age $\tau_c$$\sim$470~yr (Hu et al. 2020). 

We report here on our \textit{Chandra} Director's Discretionary Time (DDT) observation taking advantage of {\it Chandra}'s sub-arcsecond spatial resolution to study J1818 and the effect of magnetar-like outburst on its surrounding. 

\section{Observation and Data Reduction}
\label{2}

The {\it Chandra} X-ray Observatory observed J1818 with the Advanced CCD Imaging Spectrometer spectroscopic array (ACIS-S) on 2020 April 3 for an on-source exposure time of 30 ks (ObsID 23209). The source was positioned on the back-illuminated S3 chip and the data were taken in full-frame timed-exposure mode with VFAINT telemetry format. The standard processing of the data was performed using the \textit{chandra\_repro} script in CIAO version 4.12\footnote{http://cxc.harvard.edu/ciao} (CALDB 4.9.1). The event files were reprocessed (from level 1 to level 2) to remove pixel randomization and to correct for CCD charge transfer efficiencies.  An examination of the background light curves did not show any strong flares. The bad grades were filtered out and good time intervals were reserved. The resulting effective exposure after data processing was 27~ks.

\section{Imaging analysis}
\label{3.1}

Figure 1 shows broadband (0.5--7 keV) {\it Chandra} smoothed (left) and unsmoothed (right) images of J1818, with a high-energy cutoff of 7 keV since we are interested in searching for faint extended diffuse emission, and the particle background dominates this emission above $\sim$7~keV. The images are exposure-corrected using the CIAO task \textit{fluximage} with a binsize of 1~pixel. The broadband image on the left is smoothed using a Gaussian function of radius 2~pixels. We applied {\it wavdetect} tool to the ACIS-S3 cleaned image and find an X-ray source with the centre of the brightest pixel positioned at $\alpha_{J2000}$=18$^{h}$18$^{m}$00\fs23 and $\delta_{J2000}$=$-$16\degr07\arcmin52\farcs86. The uncertainty of this position is dominated by the CXO absolute position uncertainty of 0\farcs8 (at 90\% confidence level)\footnote{https://cxc.harvard.edu/cal/ASPECT/celmon/}. The image also shows evidence of diffuse emission of size $\sim$15$\arcsec$ around the magnetar.

In order to model the extent and nature of this diffuse emission, we simulated a set of 100 observations of J1818 with the \textit{Chandra} Ray Tracer (ChaRT\footnote{http://cxc.harvard.edu/ciao/PSFs/chart2/index.html}) and MARX\footnote{http://space.mit.edu/CXC/MARX/} (ver. 5.3.2) software packages, based on the parameters of the {\it Chandra} observation and blackbody spectrum as determined from the spectral fits (see Section 4). The ChaRT output was then supplied to MARX to produce the simulated event files and PSF images. Different values (0.25, 0.30, and 0.35) of the \textit{AspectBlur} parameter (which accounts for the known uncertainty in the determination of the aspect solution) were used to search for an excess corresponding to the extended emission. We created broadband (0.5--7.0 keV) radial profiles up to 15$\arcsec$ by extracting net counts in circular annuli centered on the magnetar, with an annular background region extending from 30$\arcsec$--40$\arcsec$, and rebinned the data to obtain better statistical precision. Figure~2 (left) shows the surface brightness profiles for J1818 with the profiles generated using ChaRT/MARX for different blur values. The solid horizontal line in black represents the background level. The figure shows a clear deviation from the model PSF at radii $\gtrsim$1\farcs5, and the presence of extended diffuse emission out to $\sim$10$\arcsec$ beyond which the emission becomes comparable to the background level. 

The extended emission indicates the possibility of a very compact PWN and/or a dust scattering halo. Therefore, we further examined the morphology for any evidence of asymmetry, as would be expected from a PWN. For this, we divided the 15$\arcsec$ region around J1818 into four quadrants of eight equal annuli, as shown in Figure 1 (right). The quadrants, centered on the pulsar, are oriented to the north, east, south, and west. The corresponding radial profiles are shown in Figure~2 (right). The surface brightness of the four quadrants do not vary significantly and remains consistent within errors.

\begin{figure*}[th]
\center
\includegraphics[width=0.49\textwidth]{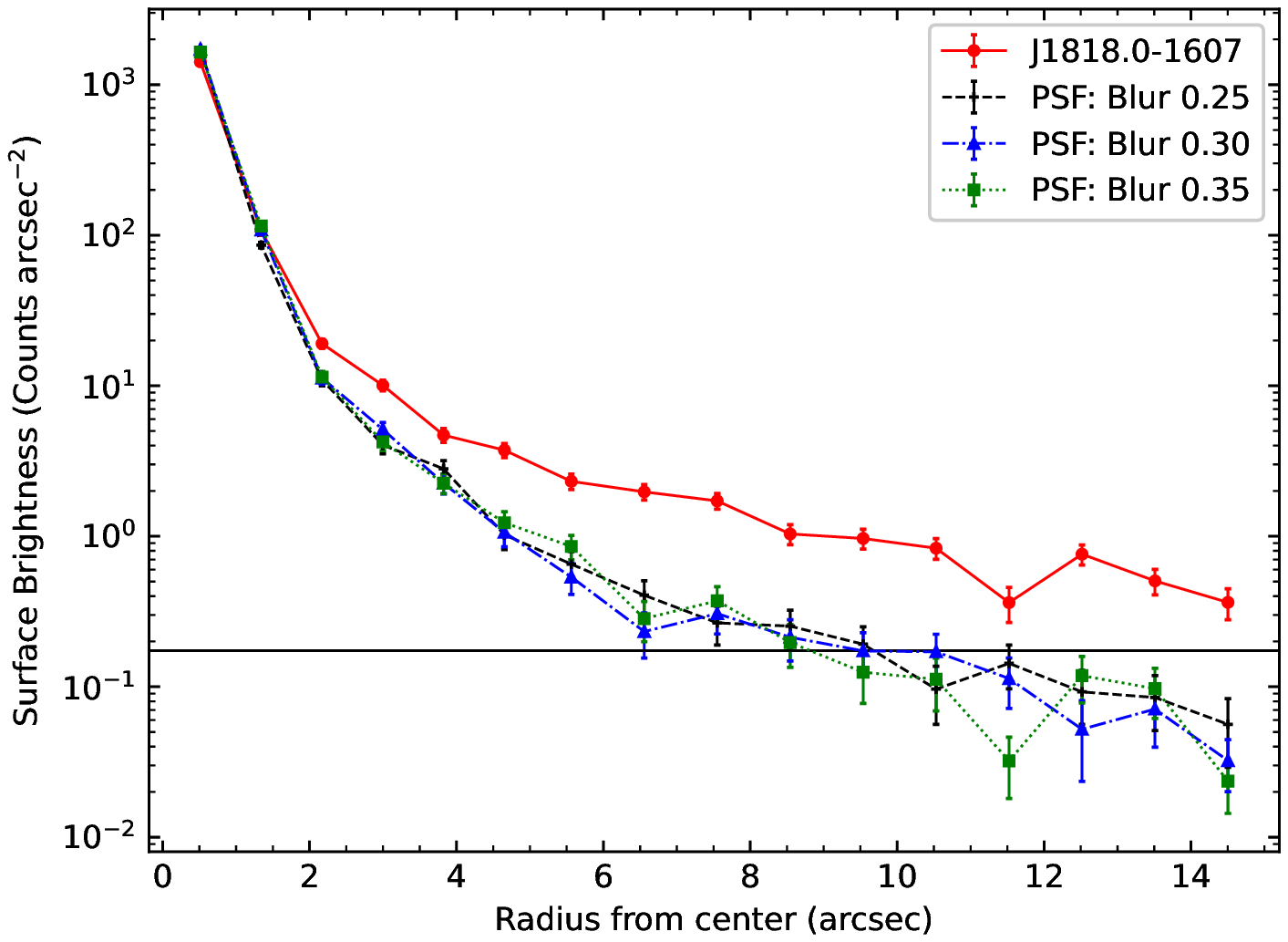}
\includegraphics[width=0.49\textwidth]{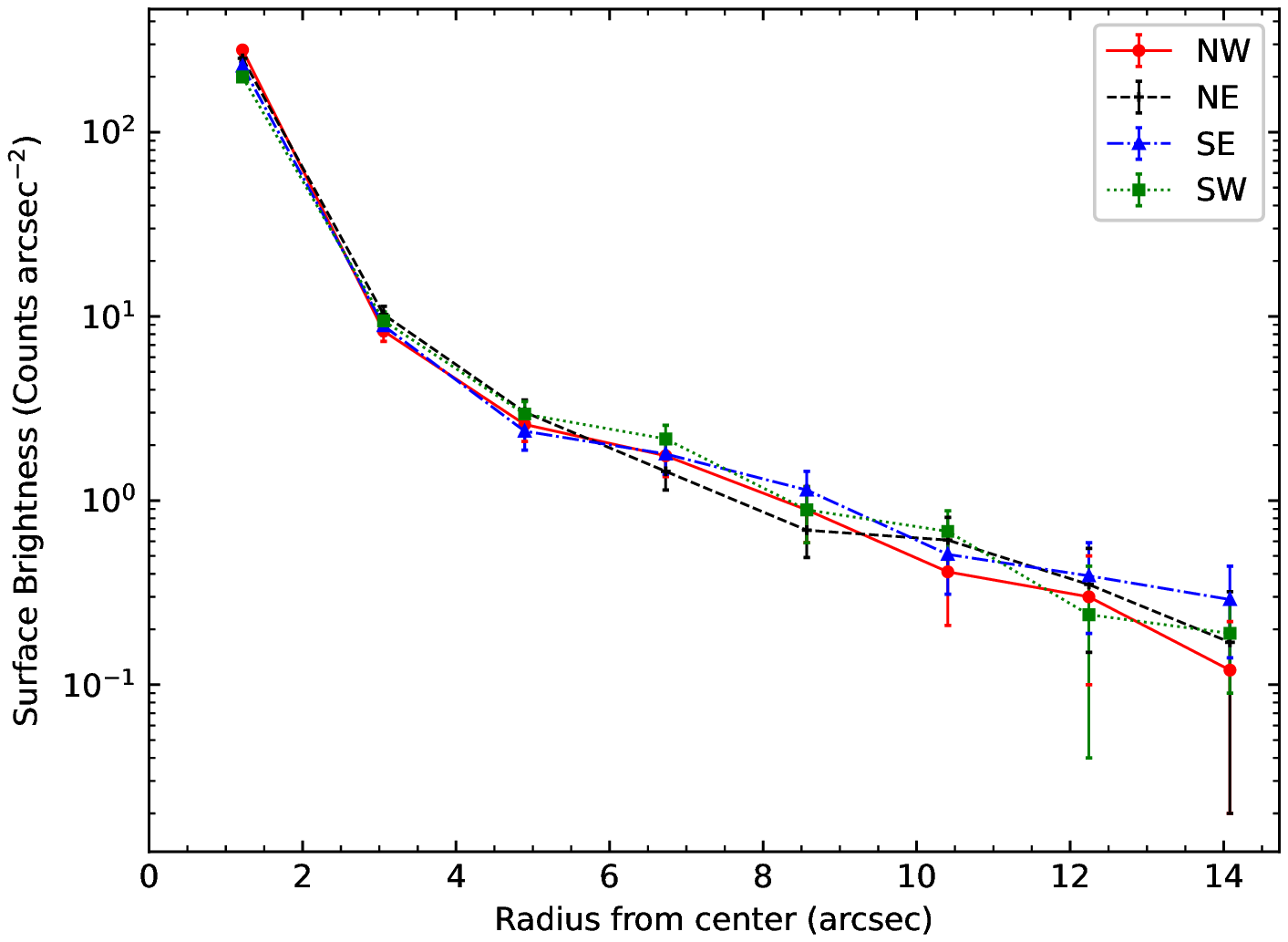}
\caption{{\it Left}: Surface brightness profiles of J1818 for different blur values. The black horizontal line represents the mean background level. {\it Right}: Radial profiles for the four quadrants shown in Figure~1 (right): north-west (NW), north-east (NE), south-west (SW), and south-east (SE). See Section 3 for details.}
\end{figure*}

\begin{figure}[ht]
\includegraphics[angle=-90, width=0.5\textwidth]{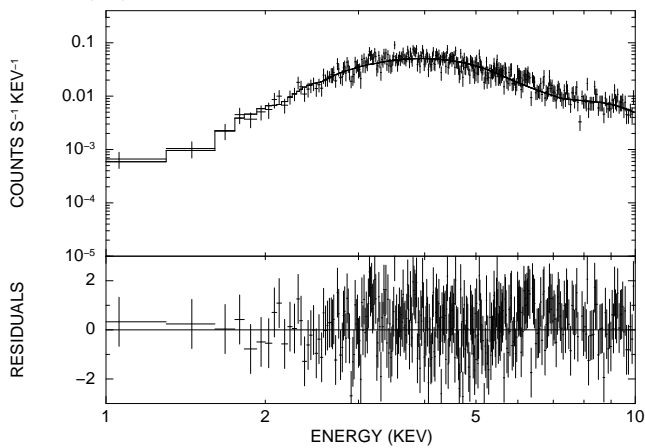}
\caption{{Left}: X-ray spectrum of Swift J1818 described by an absorbed blackbody model.   The lower panel shows residuals in units of sigmas with error bars of size one.} 
\end{figure}

\section{Spectral analysis}
\label{4}

The spectral analysis was performed with XSPEC (v12.10.1f). The contributions from background point sources were removed prior to the extraction of spectra.  All the spectra extracted were grouped by a minimum of 20 counts per bin and the errors were calculated at the 90\% confidence level. We used the $tbabs$ model (Wilms et al. 2000) to describe photoelectric absorption by interstellar medium.

The spectrum of J1818 was extracted from a 1\farcs5 radius circular region centred on the source, which encompasses more than $\sim$90\% of the encircled energy radius for a point source observed on-axis with \textit{Chandra}\footnote{http://cxc.harvard.edu/proposer/POG/html/chap6.html} at 1.49 keV.  The background was chosen from an annular ring of 3$\arcsec$--5$\arcsec$ centered on the source. As the magnetar was very bright at the time of outburst, we investigated the possibility of pileup using WebPIMMS (version 4.10) and \textit{jdpileup} model of the \textit{Chandra} spectral fitting software \textit{Sherpa} convolved with an absorbed BB model to fit the magnetar spectrum (see Section 4). The WebPIMMS and {\it Sherpa} gave a pileup fraction of 24\% and 35\%, respectively. 

\begin{table}[ht]
\caption{Spectral fits to J1818 and diffuse emission}
\begin{tabular}{l l l l l}
\hline\hline
Parameter & BB & Diffuse \\
& & emission\\
\hline
$ N_{H}$ (cm$^{-2}$) & (1.1$\pm$0.1)$\times$10$^{23}$ & 1.3$_{-0.3}^{+0.4}$$\times$10$^{23}$\\

$\Gamma$ & \nodata & 2.0$\pm$0.5  \\

kT (keV) &  1.2$\pm$0.1 & \nodata  \\

$F_{PL}^a$ & \nodata & 1.6$_{-1.0}^{+3.1}\times$10$^{-12}$  \\

$F_{BB}^a$ & 1.9$_{-0.3}^{+0.4}\times$10$^{-11}$ & \nodata &  \\

$\chi_{\nu}^2$/dof & 1.109/199 & 0.996/30  \\

$L_{X}^b$ & 9.6$_{-1.5}^{+2.0}$$\times$10$^{34}$ & 8.1$_{-0.1}^{+0.2}$$\times$10$^{33}$   \\
\hline
\end{tabular}
\tablecomments{Galactic absorption is modeled with \textit{tbabs} in XSPEC using the abundances from Wilms et al. (2000). Errors are at 90\% confidence level. \\
$^a$ Unabsorbed flux in units of ergs cm$^{-2}$ s$^{-1}$. \\
$^b$ X-ray luminosity (1--10 keV) in units of ergs s$^{-1}$ assuming isotropic emission at a distance of 6.5~kpc. \\
 }
\end{table}
 
The magnetar spectrum was fit with different models leaving all parameters to vary and included a pileup model (Davis et al. 2001) as implemented in XSPEC.  For the pileup component, only the grade-migration parameter ($\alpha$) and the fraction of events in the source extraction region within the piled-up central portion of the PSF ($psffrac$) were allowed to vary. The pileup component improved the spectral fits quality and shape of the residuals substantially. A BB model yielded a good fit with hydrogen column density $N_H$=1.1$^{+0.1}_{-0.1}$$\times$10$^{23}$~cm$^{-2}$, temperature $kT$=1.2$\pm$0.1 keV, and unabsorbed flux $F_{BB}$=1.9$^{+0.4}_{-0.3}$$\times$10$^{-11}$~ergs~cm$^{-2}$~s$^{-1}$, as summarized in Table~1. When a PL model was used ($\chi^2_{\nu}$/dof=1.075/199), the best fit values were $N_H$=1.8$^{+0.7}_{0.5}$$\times$10$^{23}$~cm$^{-2}$ and photon index $\Gamma$=2.2$\pm$0.3 with an unabsorbed flux $F_{PL}$=4.5$^{+1.0}_{-0.8}$$\times$10$^{-11}$~ergs~cm$^{-2}$~s$^{-1}$. Although the PL model gave comparable fit statistics as the BB model, we obtained a flux exceeding the source flux at the time of outburst (see Esposito et. al. 2020), likely due to high pileup. Since there is also little PL contribution below 7 keV, we conclude that a single absorbed BB provides an adequate model to the source spectrum. The spectral fits were explored with different background regions and we found no significant differences in the spectral parameters. The addition of a second component was statistically not required.  The best-fit BB spectrum is shown in Figure 2.

We determined the physical properties of the diffuse emission (see Figure~1) by selecting an annular ring of 3$\arcsec$--10$\arcsec$ region and a background of 30$\arcsec$--40$\arcsec$ radius centered on J1818. The extended emission has a total of 532$\pm$25 background subtracted counts in the 1--10 keV range. An absorbed PL model provided a good fit with $N_H$=1.3$^{+0.4}_{-0.3}$$\times$10$^{23}$~cm$^{-2}$, $\Gamma$=2.0$\pm$0.5, and an unabsorbed flux of 1.6$^{+3.1}_{-1.0}$$\times$10$^{-12}$~ergs~cm$^{-2}$~s$^{-1}$, as shown in Table~1.

\section{Discussion and Conclusion}
\label{5}

In this section, we discuss the analysis of {\it Chandra} observations of the newly discovered magnetar Swift J1818.0--1607 and associated compact extended emission, and explore its environment. The distance to the magnetar is estimated to be in the range of 4.8--8.1 kpc based on the Cordes-Lazio Galactic free electron density (NE2001; Cordes \& Lazio 2002) and YMW2016 (Yao et al. 2017) models. We adopt an average distance of 6.5~kpc and introduce a scaling factor $d_{6.5}$=$d$/6.5 kpc to account for the distance uncertainty. 

J1818 is the fifth magnetar to show radio emission and detected simultaneously in X-rays. It is also among the youngest magnetars in the Galaxy with an inferred characteristic age $\sim$470 yrs (Hu et al. 2020). If the true age of J1818 is comparable to its characteristic age, we expect a young supernova remnant (SNR) surrounding the magnetar. The detection of any associated SNR shell will also help constrain the pulsar's true age since the characteristic age is unreliable, largely due to the implicit assumptions that all pulsars are born spinning very fast and spinning down under a constant braking index of 3. Therefore, we investigate the environment of J1818 by constructing a composite colour image using the radio (red), infrared (green), and X-ray (blue) data, as shown in Figure~4. The radio image was obtained from the Multi-Array Galactic Plane Imaging Survey (MAGPIS\footnote{http://third.ucllnl.org/gps/}) at a wavelength of 20~cm (Helfand et al. 2006) and the infrared image from the survey of the inner Galactic plane using \textit{Spitzer}'s Multiband Imaging Photometer at a wavelength of 24~$\mu$m (MIPSGAL\footnote{http://mipsgal.ipac.caltech.edu}; Carey et al. 2009). The infrared image shows that the magnetar lies in a complex region of the Galactic plane with several foreground or background sources, appearing in green. 

To the northwest of the magnetar lies an HII region, G014.576+0.091 as identified in the WISE catalogue\footnote{http://astro.phys.wvu.edu/wise/}, at $l$=14.467, $b$=0.091 and has a size of $\sim$425$\arcsec$.  The near and far distance estimates to the HII region are 3.7~kpc and 12.8~kpc (Table~6 of Anderson et al. 2014). We notice a nearly complete shell-like structure showing 20 cm continuum emission (in red) overlapping with this HII region. The 24~$\mu$m emission (in green) is also seen from the shell, but not inside as has been observed for HII regions.  However, there are no catalogued SNRs around this position (Green 2019; Ferrand \& Safi-Harb 2012\footnote{http://snrcat.physics.umanitoba.ca}). Therefore we speculate that this shell may be a possible candidate SNR at $l$=14.57, $b$=0.14, with a size of $\sim$11$\arcmin$.  The position of J1818 is roughly 20$\arcmin$ from the candidate SNR's centre.  If we assume that J1818 was born at the centre of this candidate SNR, and further assume an SNR age of $\sim$5--10 kyr (which is typical for SNRs in the Sedov phase), the magnetar would need a projected velocity of $\sim$3600--7300 km~s$^{-1}$ to reach its current location at a distance of 6.5 kpc. This estimate would be even higher if the SNR candidate were younger, making the association very unlikely. Proper motion measurements of J1818 and high resolution radio observations are required to investigate any possible association, as well as confirm the nature and extent of the shell. 

\begin{figure*}[htp]
\includegraphics[width=\textwidth]{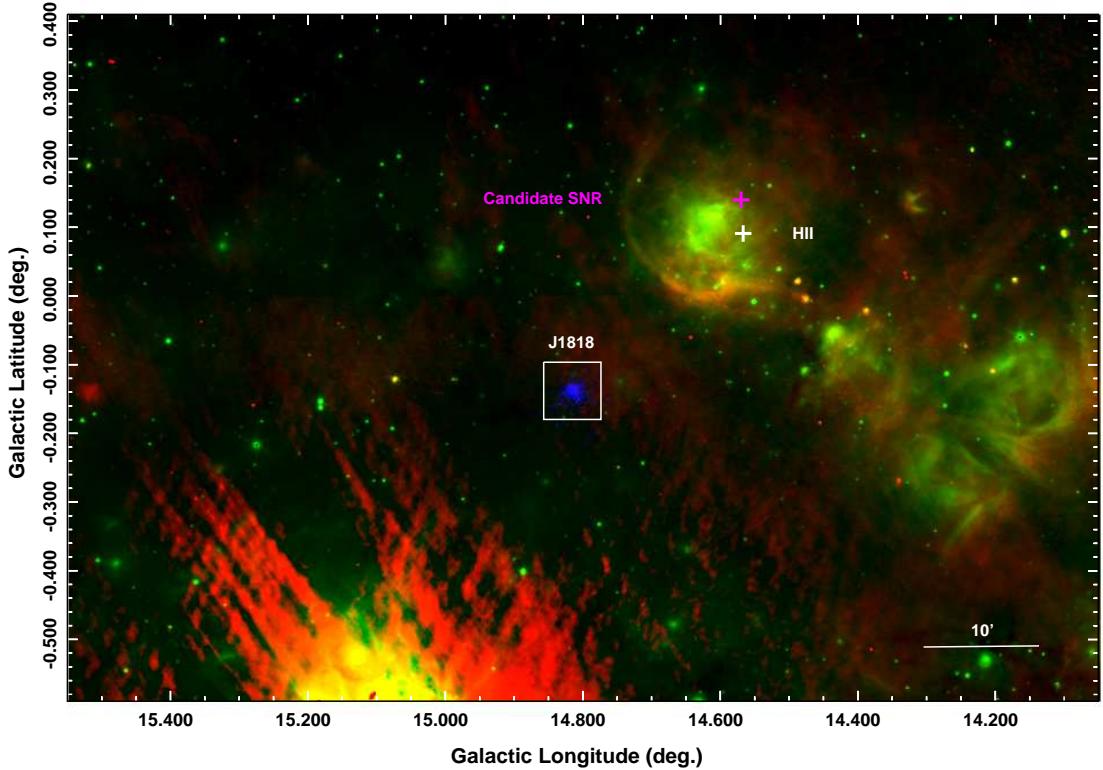}
\caption{Multiwavelength image of the environment of J1818 with the MAGPIS 20 cm radio in red, MIPSGAL 24$\mu$m infrared in green, and 0.5--7 keV {\it Chandra} image in blue. The images have been smoothed using a Gaussian function of radius 3 pixels. To the northwest of the magnetar lies an HII region overlapping with a candidate SNR, marked by magenta and white crosses, respectively. The centres of HII region and candidate SNR are marked by magenta and white crosses, respectively.  The extremely bright source to the southeast is M17 (Messier 17). North is up and East is to the left. See Section~5 for details.}
\end{figure*}

The \textit{Chandra} spectrum of J1818 is best described by a BB model with $kT$=1.2$\pm$0.1~keV and an unabsorbed flux of 1.9$_{-0.3}^{+0.4}\times$10$^{-11}$~ergs~cm$^{-2}$~s$^{-1}$. Assuming isotropic emission, the magnetar's X-ray luminosity is $L_{X}$=9.6$_{-1.5}^{+2.0}$$\times$10$^{34}$~$d_{6.5}^2$~ergs~s$^{-1}$, implying an efficiency $\eta_{X}$=$L_{X}/\dot{E}$$\sim$0.13~$d_{6.5}^2$ in the 1--10 keV energy range. The neutron star's emitting radius inferred from the BB fit is 0.6$\pm$0.1~km, slightly different from those reported by Esposito et. al (2020) and Hu et al. (2020). This can be attributed to {\it Chandra}'s high imaging resolution which allows us to disentangle the diffuse emission from the point source, as opposed to {\it XMM-Newton} or NICER. In the twisted magnetosphere model (Thompson et al. 2002), the thermal emission of magnetars is thought to originate from heating within the star due to the decay of the strong internal magnetic field.  Twists in the magnetosphere suggest external heating to the surface. Twisted magnetic fields allow for the development of electric fields, accelerating particles off the surface resulting in a Comptonized BB-like spectrum (Beloborodov 2009). These particles can return to the surface, heating it through particle bombardment. For internal heating, it is simple heat diffusion towards the surface from the hotter core. Comparing the results obtained for J1818 with those of magnetars and high-$B$ pulsars, its X-ray properties are similar to those seen in transient magnetars. Following outbursts, transient magnetars typically have high $kT$ ($>$0.7~keV) compared to their quiescent BB temperatures of $\approx$0.4~keV (Coti Zelati et al. 2017). It is also worth mentioning that J1818 has the highest spin-down luminosity (7.2$\times$10$^{35}$~ergs~s$^{-1}$) among the magnetars, followed by the radio magnetar 1E 1547.0--5408 with $\dot{E}$=2.1$\times$10$^{35}$~ergs~s$^{-1}$.  However, the inferred X-ray efficiency of J1818 is more in line with those of the high-$B$ pulsars showing magnetar-like bursts with $L_X/\dot{E}<<$1. This, and the relatively high $\dot{E}$, imply that J1818 could be at least partly powered by rotation, similar to the rotation-powered pulsars. 

The imaging analysis resolved a compact diffuse emission of $\geq$10$\arcsec$ radius around J1818, with $N_H$=1.3$_{-0.3}^{+0.4}\times$10$^{23}$~cm$^{-2}$, hard $\Gamma$=2.0$\pm$0.5, unabsorbed flux of 1.6$_{-1.0}^{+3.1}\times$10$^{-12}$~ergs~cm$^{-2}$~s$^{-1}$, and a luminosity of 8.1$^{-0.1}_{+0.2}\times$10$^{33}$~$d_{6.5}^2$~ergs~s$^{-1}$ in the 1--10 keV band. The heavy absorption towards this region, combined with the brightening of the magnetar and its location in a crowded region of the Galactic plane, should cause scattering of point source radiation by dust along the line of sight forming a dust scattering halo. Such halos have been observed around high-$B$ pulsars, or magnetars (during an outburst), with index ranging from $\sim$2 to 5 as reported by several authors (e.g.,  Gotthelf et al. 2020; Tiengo et al. 2010; Younes et al. 2012; Israel et al. 2016).  A dust scattering halo typically shows symmetric structures around the point source and a softer spectrum than the illuminating source (i.e., the magnetar), since the scattering cross-section of the dust particles scales as $E^{-2}$ of the incident photon energies (Rivera-Ingraham \& van Kerkwijk 2010). This is indeed the case here, where the diffuse emission seems fairly symmetric from the radial profile analysis and its spectrum marginally softer than J1818 (see Section~4). The magnetar spectrum below 10 keV is fit with a hotter BB temperature (Table~1) and a hard X-ray tail with $\Gamma$=0.0$\pm$1.3  in the 1--20 keV band 3 days post outburst (Esposito et al. 2000). Both of these components are very hard and can result in a halo with a flatter spectrum (although softer than the source) as seen here with {\it Chandra}. Hence, it is likely that the diffuse emission around J1818 is dominated by a dust scattering halo associated with the magnetar burst. Esposito et al. (2020) had also reported diffuse emission at radial distances of 50\arcsec--110\arcsec\ from J1818, which could be the halo that was brighter at the time of {\it XMM-Newton} observations. Since the {\it Chandra} observation was taken 21 days post outburst, it is possible that the halo has already become dimmer and smaller in size.

X-ray PWNe, with $\Gamma$=1--2.5, have been observed around young ($\tau_c$$\sim$0.6--30 kyr) rotation-powered pulsars (RPPs) with spin-down power $\dot{E}$ ranging from $\sim$10$^{33}$--10$^{38}$ ergs~s$^{-1}$ (Kargaltsev \& Pavlov 2008; Kargaltsev et al. 2013). J1818's young inferred age and relatively high $\dot{E}$ with respect to magnetars imply that it could power a compact PWN. The position of J1818 was observed previously with {\it Chandra} (ObsID: 8160, 2.7~ks exposure) in 2008 and {\it XMM-Newton} (ObsID: 0800910101, 60~ks exposure) in 2018. However, no point source or nebula emission was detected in these observations. It is possible that a faint nebula could be associated with J1818, but not detected in previous observations given the lack of sensitivity to such compact diffuse emission combined with heavy absorption. We can estimate an upper limit on the flux of a possible PWN undetected in previous {\it Chandra} observation (see Esposito et al. 2020 for {\it XMM-Newton} data) for the same extraction regions as in Section~4 with the ciao tool {\it srcflux}. We assume an $N_H$=1.3$\times$10$^{23}$ cm$^{-2}$ and a $\Gamma$=2, as is typical of PWNe. The 3$\sigma$ upper limit on the unabsorbed model flux is $\sim$2.3$\times$10$^{-13}$ ergs~cm$^{-2}$~s$^{-1}$, corresponding to a luminosity of $\sim$1.2$\times$10$^{33}$~$d_{6.5}^2$~ergs~s$^{-1}$ and efficiency $\sim$0.002~$d_{6.5}^2$. This efficiency falls within the expected range of $\sim$10$^{-6}$--10$^{-1}$ observed in young pulsars with PWNe (Kargaltsev \& Pavlov 2008) and is comparable to that of the high-$B$ pulsar J1119--6127 which showed a magnetar-like burst in 2016. It is, however, lower than what is typically observed in magnetars (see Table~1 of Safi-Harb 2013), suggesting that J1818 behaves more like a high-$B$ rotation-powered pulsar than a magnetar.  Interestingly, J1119--6127 also hosts a compact ($\sim$10$\arcsec$ in radius) nebula with a quiescent luminosity of $\sim$1.9$\times$10$^{32}$~ergs~s$^{-1}$ (Safi-Harb \& Kumar 2008; Blumer et al. 2017). The nebula was first detected in a 60 ks {\it Chandra} observation (Gonzalez \& Safi-Harb 2003) and could not be resolved with \textit{XMM-Newton}. Hence, deep \textit{Chandra} observations of the magnetar in quiescence are necessary to investigate the nature of this extended emission and to search for the existence of any possible compact wind nebula.

To conclude, the sensitivity and resolution offered by {\it Chandra} allowed us to study the recently discovered magnetar Swift J1818.0--1607 and resolve compact faint emission surrounding it. Our study points to J1818 being a transient source showing properties intermediate between high-$B$ pulsars and magnetars, and to the diffuse emission being dominated by a dust scattering halo. Future deep \textit{Chandra} observations of the source in quiescence will confirm the nature of this extended emission and place further constraints on any underlying compact wind nebula powered by the rotational energy loss of the pulsar.

\acknowledgments
The authors thank \textit{Chandra} science team for making this DDT observation possible, and Loren D. Anderson for useful discussions and providing the high-resolution MAGPIS and SPITZER images. We also thank the anonymous referee for comments that helped improve the manuscript. Support for this work was provided by the NASA grant DD0-21116X. SSH acknowledges support by the Natural Sciences and Engineering Research Council of Canada and the Canadian Space Agency.

\end{document}